\documentclass{llncs}
\usepackage[latin1]{inputenc}
\usepackage{amssymb,amsmath,array}
\usepackage{url}
\usepackage{graphicx}
\usepackage{dsfont}
\usepackage{xcolor}
\usepackage{subfigure}
\usepackage[english]{babel}

\graphicspath{{images/}}

\begin{document}
\title{Study of a bias in the offline evaluation of a recommendation algorithm}

\author{Arnaud de Myttenaere\inst{1,2} \and Boris Golden\inst{1} \and B\'en\'edicte Le Grand\inst{2} \and Fabrice Rossi\inst{2}}
\institute{Viadeo \and Universit\'e Paris 1 Panth\'eon - Sorbonne}


\maketitle
\begin{abstract}
Recommendation systems have been integrated into the majority of large online
systems to filter and rank information according to user profiles. It thus
influences the way users interact with the system and, as a consequence,
bias the evaluation of the performance of a recommendation algorithm computed using
historical data (via \emph{offline evaluation}). This paper describes this bias and discuss the relevance of a weighted offline evaluation to reduce this bias for different classes of recommendation algorithms.
\end{abstract}

\section{Introduction}
A recommender system provides a user with a set of possibly ranked items that
are supposed to match the interests of the user at a given moment
\cite{park2012literature,kantor2011recommender,adomavicius2005toward}. Such
systems are ubiquitous in the daily experience of users of online systems. For
instance, they are a crucial part of e-commerce where they help consumers
select movies, books, music, etc. that match their tastes. They also
provide an important source of revenues, e.g. via targeted ad placements
where the ads displayed on a website are chosen according to the user profile
as inferred by her browsing history for instance. Commercial aspects set
aside, recommender systems can be seen as a way to select and sort information
in a personalized way, and as a consequence to adapt a system to a user.

Obviously, recommendation algorithms must be evaluated before and during their
active use in order to ensure their performance. Live monitoring is generally achieved using online performance metrics (e.g. click-through rate of displayed ads) and several recommendation strategies can be compared using AB testing and online evaluation, whereas offline evaluation is computed using historical data. However putting an algorithm in production, collect and analyze data is generally a long process (many days or weeks). Offline evaluation allows to quickly test several strategies without having to wait for real metrics to be collected nor impacting the performance of the online system. One of the main strategy of offline evaluation consists in simulating a recommendation by removing a confirmation action (click, purchase, etc.)  from a user profile and testing whether the item associated to this action would have been recommended based on the rest of the profile \cite{shani2011evaluating}. Numerous variations of this general scheme are used ranging from removing several confirmations to taking into account item ratings.

While this general scheme is completely valid from a statistical point of
view, it ignores various factors that have influenced historical data as the recommendation algorithms previously used. Even if limits of evaluation strategies for recommendation algorithms have been identified (\cite{HerlockerEtAl2004Evaluating,mcnee2006being,said2013user}), this protocol is still intensively used in practice.

We study in this paper the general principle of instance weighting proposed in \cite{demytt2014reducing} and show its practical relevance on the simple case of constant recommendation and on two collaborative filtering algorithms. In addition to its good performances, this method is more realistic than solutions proposed in \cite{HerlockerEtAl2004Evaluating,mcnee2006being} for which a data collection phase based on random recommendations has to be performed. While this phase allows one to build a bias free evaluation data set, it has also adverse effects in terms of e.g. public image or business performance when used on a live system, as random recommendations are obviously less relevant than personnalized recommendations got by an algorithm.

The rest of the paper is organized as follows. Section \ref{sec:problem-formulation} describes in details the setting and the problem. Section \ref{sec:reduc-eval-bias} introduces the
weighting scheme proposed to reduce the evaluation bias. Section
\ref{sec:exper-eval-constant} demonstrates the practical relevance of our method for the particular case of constant algorithms and present experimental results based on real world data extracted from Viadeo (professional social
network\footnote{See \url{http://corporate.viadeo.com/en/} for more information about Viadeo.}). Section \ref{sec:exper-eval-cf} describes the results of our approach on two collaborative filtering and discuss the reduction of the bias for elaborated algorithms.

\section{Problem formulation}\label{sec:problem-formulation}

\subsection{Notations}
\begin{figure}[htb]
  \centering
\includegraphics[width=3.5cm]{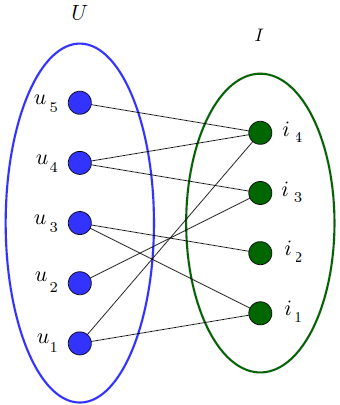}
  \caption{Representation of the data as a bipartite graph and notations}
  \label{fig:schemaData}
\end{figure}

We denote $U$ the set of users, $I$ the set of items and $\mathcal{D}_t$ the historical data available at time $t$. As user are associated to items, $\mathcal{D}_t$ can be represented as a bipartite graph. Let $n_U = \#U$ and $n_I = \#I$ be the cardinal of the set of users and items at time $t$, and $A$ represents the adjacency matrix of the bipartite graph given by $\mathcal{D}_t$. Then $A = \left( \begin{matrix}0_{n_I} & B \\ B^T & 0_{n_U} \end{matrix}\right)$ where $0_{n}$ represents the zero matrix of size $n \times n$, and $B$ is a $n_I \times n_U$ binary matrix. $B$ is called biadjacency matrix and for each $(i, u)$ in $I \times U$, $b_{i,u} = 1$ if the item $i$ is associated to user $u$ and 0 else. A representation of the data is presented on figure~\ref{fig:schemaData}

A recommendation algorithm is a function $g$ from $U\times \mathcal{D}_t$ to some set built from $I$. We will denote $g_t(u) = g(u,\mathcal{D}_t)$ the recommendation computed by the algorithm $g$ at instant $t$ for user $u$.
We assume given a quality function $l$ from the product of the result space of $g$ and $I$ to $\mathbb{R}^+$ that measures to what extent an item $i$ is correctly recommended by $g$ at time $t$ via $l(g_t(u),i)$. We denote $I_u$ the items associated to a user $u$, and $U_i$ the set of users which are associated to the item $i$.

\subsection{The classical offline evaluation procedure}\label{sec:offlineProcedure}
Offline evaluation is based on the possibility of ``removing'' any item $i$
from a user profile, which can be computed using stochastic or exhaustive sampling. Although exhaustive sampling gives more robust results, the stochastic approach is often prefered (especially for large systems) as it is faster and often precise enough to compare several algorithms. The user profile got after removing item $i$ from user $u$ is denoted $u_{-i}$ and $g_t(u_{-i})$ is the recommendation obtained at instant $t$ when $i$ has been removed from the profile of user $u$.

Finally, offline evaluation follows a general scheme in which a user is chosen
according to some probability on users $P(u)$, which might reflect the business importance of the users. Given a user, an item $i$ is chosen among the items associated to its profile, according to some
conditional probability on items $P(i|u)$. When an item $i$ is not associated
to a user $u$ (that is $i\not\in I_u$), $P(i|u)=0$. A very common choice for $P(u)$ is the uniform probability on $U$ and it is also very common to use a uniform probability for $P(i|u)$ (other strategy could favor items recently associated to a profile). As the system evolves over the time, $P(u)$ and $P(i)$ depends on $t$.

The two distributions $P(u)$ and $P(i|u)$ lead to a joint distribution $P(u,i)=P(i|u)P(u)$ on $U\times I$. In other words, the classical offline evaluation consists in selecting a random node in user's part of the bipartite graph, and then a random node among the ones associated to the selected user. Many other graph sampling methods could be used (random edge selection, \dots)

\subsection{Origin of the bias in offline evaluation}
As presented in \cite{li2011unbiased,demytt2014reducing} the classical offline evaluation procedure ignores various factors that have influenced historical data as the recommendation algorithms previously used, promotional offers on some specific products, etc. Assume for instance that several recommendation algorithms are evaluated at
time $t_0$ based on historical data of the user database until $t_0$. Then the best
algorithm is selected according to a quality metric associated to the offline
procedure and put in production. It starts recommending items to the
users. Provided the algorithm is good enough, it generates some confirmation
actions. In other words, the recommendation campaigns introduce many new vertices in the bipartite graph representing the data (see figure~\ref{fig:schemaData}). Those actions can be attributed to a good user modeling but also to
luck and to a natural attraction of some users to new things. This is
especially true when the cost of confirming/accepting a recommendation is
low. In the end, the state of the system at time $t_1>t_0$ has been influenced by the recommendation algorithm in production. 
   
 Then if one wants to monitor the performance of this
algorithm at time $t_1$, the offline procedure tends to overestimate the
quality of the algorithm because confirmation actions are now frequently
triggered by the recommendations, leading to a very high predictability of the
corresponding items. 

\begin{figure}[htb]
  \centering
\includegraphics[width=7cm]{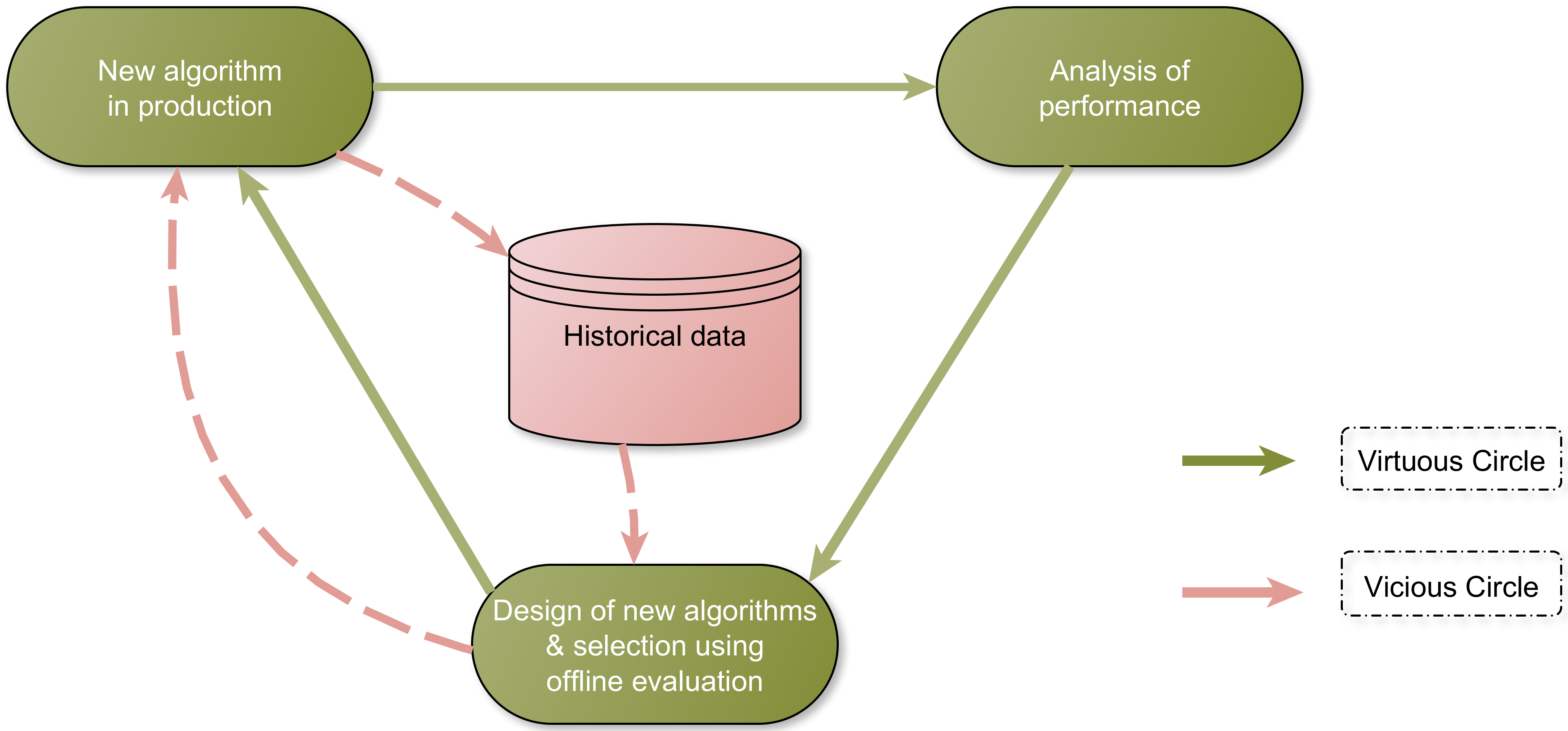}
  \caption{The evolution of the recommendation system}
  \label{fig:schema}
\end{figure}

Finally, one can decompose the evolution of a recommendation system in two cycles represented in figure~\ref{fig:schema}. On one hand there is a virtuous circle (also called \textit{lean} circle) in three steps: first an algorithm is put in production and the data collection process starts, then the collected data are analyzed to measure the performance of the algorithm, and finally data are also used to select the best algorithm among several new ones by offline evaluation. On the other hand we also observe a vicious circle as the algorithm in production influences the users behaviors, which introduces a bias in historical data used for the offline evaluation procedure.

This bias in offline evaluation with online systems can also be caused by other
events such as a promotional offer on some specific products between a first
offline evaluation and a second one. The main effect of this bias is to favor algorithms
that tend to recommend items that have been favored between $t_0$ and $t_1$
and thus to favor a kind of ``winner take all'' situation in which the
algorithm considered as the best at $t_0$ will probably remain the best one
afterwards, even if an unbiased procedure could demote it.  Indeed the score of an algorithm in production, given by the classical offline evaluation, tends to increase over time. More generally, the classical offline evaluation  tends to overestimate (resp. underestimate) the unbiased score of an algorithm similar (resp. orthogonal) to the one in production.

More formally, the classic offline evaluation procedure consists in calculating the quality of the recommendation algorithm $g$ at instant $t$ as
$L_t(g)=\mathbb{E}(l(g_t(u_{-i}),i))$ where the expectation is taken with respect
to the joint distribution:
\begin{equation}\label{eq:loss}
L_t(g)=\sum_{(u,i)\in U\times I}P_t(i|u)P_t(u)l(g_t(u_{-i}),i).
\end{equation}

 Then if two algorithms are evaluated at two different moments, their qualities are not directly comparable. Although as in an online system $P(i|u)$ evolves over time\footnote{even if $P(u)$ could also evolve over time we do not consider the effects of such evolution in the present article.}
once a recommendation algorithm is chosen based on a given state of the system, it starts influencing the state of
the system when put in production, inducing an increasing distance between its
evaluation environment (i.e. the initial state of the system) and the
evolving state of the system. This influence is responsible for a bias on offline evaluation as it relies on historical data.

A naive solution to correct this bias would be to compare algorithms only with respect to the original
database at $t_0$, but this approach is not optimal as it would discard natural evolutions of user profiles.

\subsection{Impact of recommendation campaigns on real data}\label{sec:real-world-illustr}
\begin{figure}[h]
  \centering
  \begin{subfigure}
  \centering
	\includegraphics[width=4.5cm]{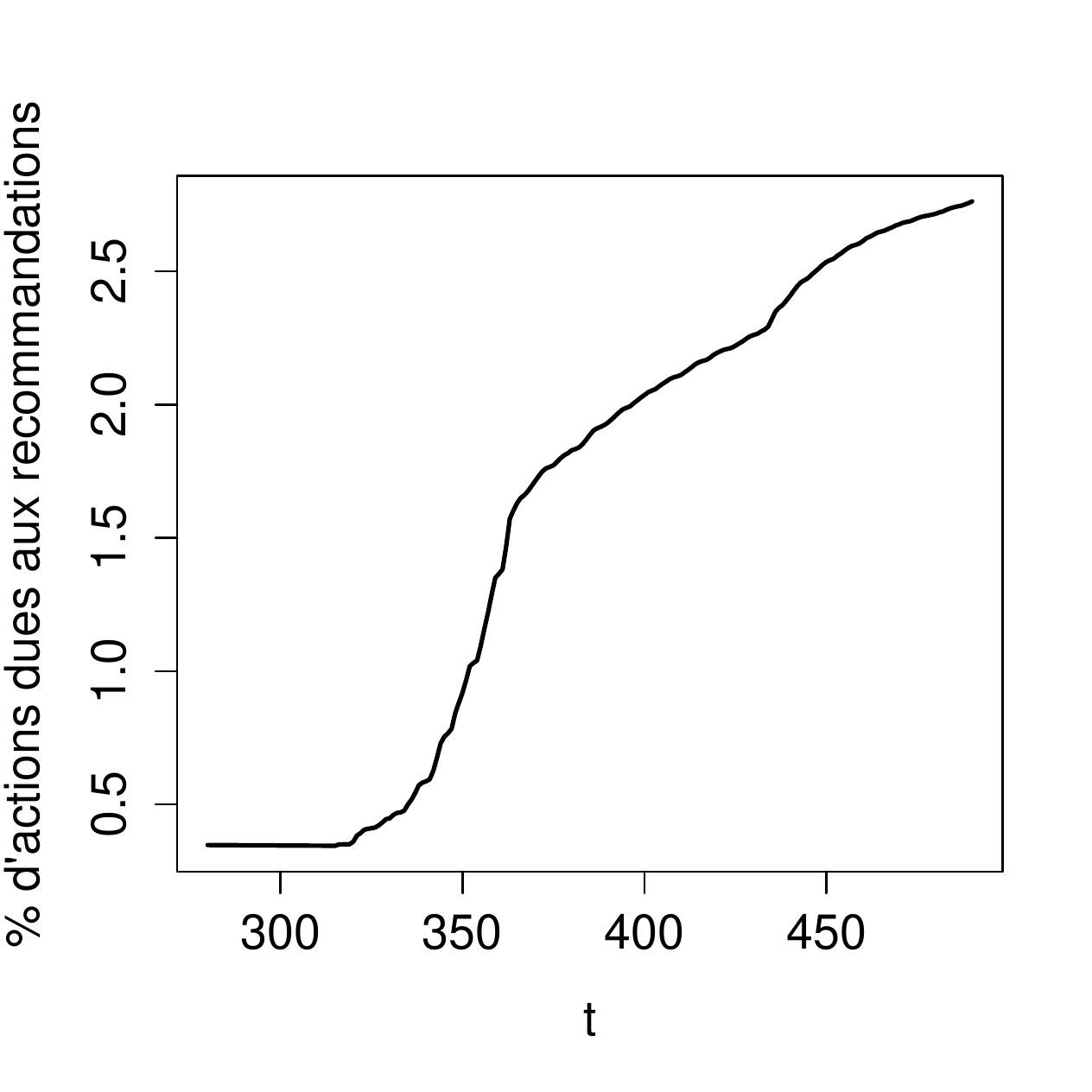}
  \end{subfigure}  
  \begin{subfigure}
  \centering
	\includegraphics[width=4.5cm]{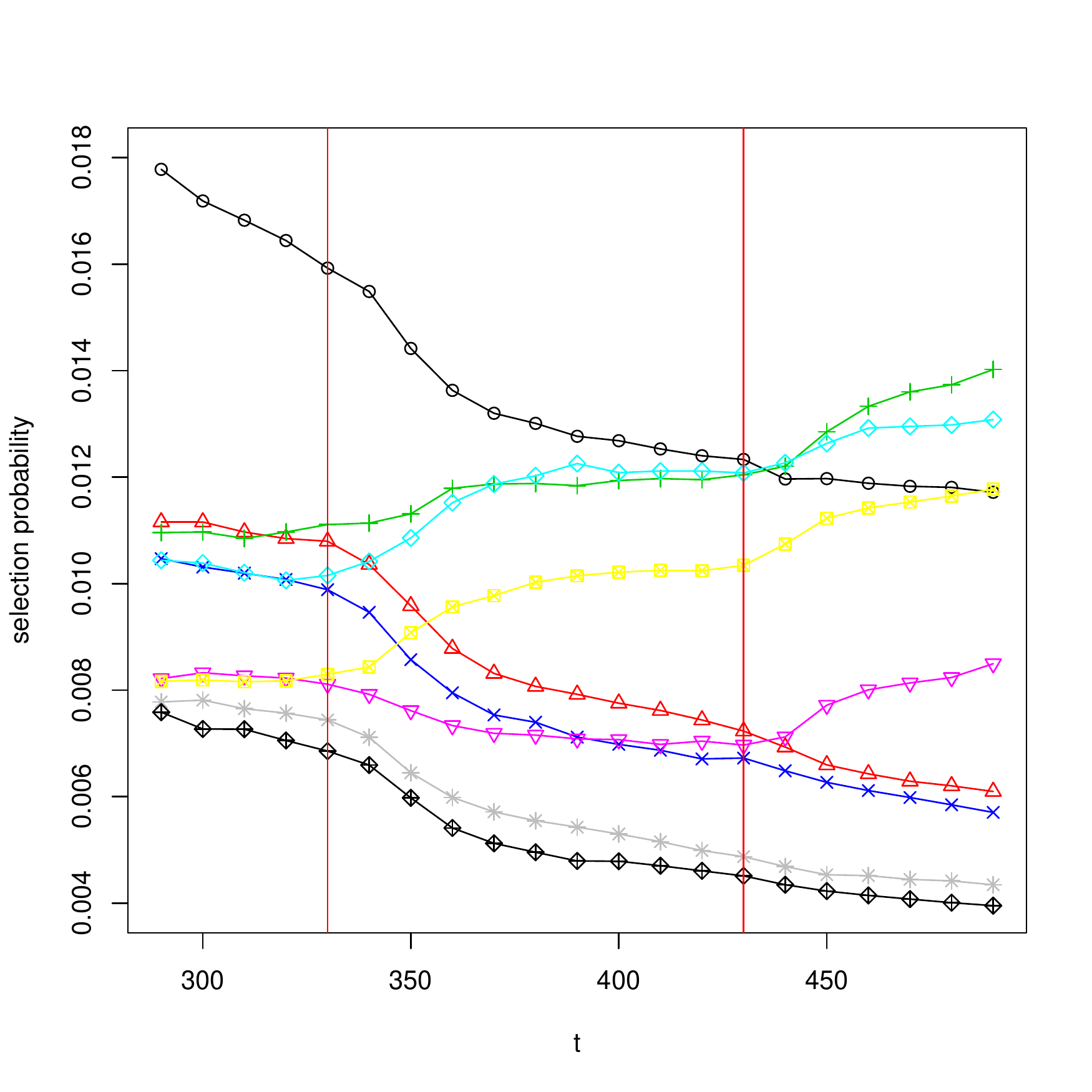}  
  \end{subfigure}
  \caption{Impact of recommendation campaigns on the item
probabilities: the left figure displays the percentage of observations induced
by the recommendations, while the right figure shows examples of the evolution
of $P(i)$ through time.}
  \label{fig:impact}
\end{figure}

We illustrate the evolution of the $P_t(i)$ probabilities in an online system with
a functionality provided by the Viadeo platform: each user can claim to have
some skills that are displayed on his/her profile (examples of skills include
project management, marketing, etc.). In order to obtain more complete profiles,
skills are recommended to the users via a recommendation algorithm, a practice
that has obviously consequences on the probabilities $P_t(i)$, as illustrated on Figure
\ref{fig:impact}. 

The skill functionality has been implemented at time $t=0$. After 300 days,
some of the $P_t(i)$ are roughly static. Probabilities of other items still
evolve over time under various influences, but the major sources of evolution
are recommendation campaigns.  Indeed, at times $t=330$ and $t=430$,
recommendation campaigns have been conducted: users have received personalized
recommendation of skills to add to their profiles. The figure shows strong
modifications of the $P_t(i)$ quickly after each campaign.  In particular, the
probabilities of the items which have been recommended increase significantly;
this is the case for the green, yellow and turquoise curves at $t=330$. On the
other hand, the probabilities of the items which have not been recommended
decrease at the same time. The probabilities tend to become stable again until
the same phenomenon can be observed right after the second recommendation
campaign at $t=430$: the curves corresponding to the items that have been
recommended again keep increasing. The purple curve represents the probability selection
of an item which has been recommended only during the second recommendation
campaign. Section \ref{sec:impactOnReco} demonstrates the effects of this
evolution on the evaluation of recommendation algorithms.

\section{Reducing the evaluation bias}\label{sec:reduc-eval-bias}

\subsection{A weighted offline evaluation method to reduce the bias}

A simple transformation of equation (\ref{eq:loss}) shows that for a constant algorithm $g$ (i.e. if recommendations are the same for every users): $\label{eq:loss:constant} L_t(g) =\sum_{i\in I}P_t(i)l(g_t,i)$. As a consequence, a way to guarantee a stationary evaluation framework for a constant algorithm is to have constant values for the marginal distribution of the items, $P_t(i)$.

A natural solution would be to record those probabilities at $t_0$ and use
them as the probability to select an item in offline evaluation at $t_1>t_0$. However, as the selection of users and items leads to a joint distribution, this would require to revert the way offline evaluation is done: first select an item, then select a user having this item with a certain probability $\pi_t(u|i)$ leading to a different probability of users selection. Finally this process lead to a similar problem on users, and as in most of systems $\#U > \#I$, it is more efficient to keep the classical evaluation protocol (see section~\ref{sec:complexity} for more details).

Moreover, we will see that the recalibration of every items is not necessary to reduce the main part of the bias. Indeed in practice most of the time a few items concentrate most of the recommendations (very popular items, discount on selected products, ...). Thus one can reduce the major part of the bias by optimizing the weight of the $p$ items such that the deviation given by $|P_{t_0}(i) - P_{t_1}(i)|$ have the highest values. In practice the choice of $p$ is done according to practical (time) or business constraints. 

Thus the weighting strategy that we described in \cite{demytt2014reducing} consists in keeping the classical choice for $P_t(u)$ and weighting $P_t(i|u)$ by departing from the classical values for $P_t(i|u)$ (such as using a uniform probability) in order to mimic static values for $P_ {t_0}(i)$ by :
\begin{equation*}
  \label{eq:weighted:conditional}
P_t(i|u,\omega)=\frac{\omega_iP_t(i|u)}{\sum_{j\in I_t}\omega_jP_t(j|u)}.
\end{equation*}

These weighted conditional probabilities lead to weighted item probabilities defined by:
\begin{equation*}
  \label{eq:weighted:item}
P_t(i|\omega)=  \sum_{u\in  U}P_t(i|u,\omega)P_t(u).
\end{equation*}

Then we suggest to minimize the distance between $P_{t_1}(i|\omega)$ and $P_{t_0}(i)$ by optimizing the Kullback-Leibler divergence, defined by :
\begin{equation}
  \label{eq:divergence}\notag
D(\omega)=\sum_{i\in I_{t_0}}P_{t_0}(i)\log\frac{P_{t_0}(i)}{P_{t_1}(i|\omega)}
\end{equation}
where $I_{t_0}$ represents the set of items present at $t_0$. The asymmetric nature of this distance is useful in our context to consider time $t_0$ as a reference. Moreover this asymmetry reduces the influence of rare items at time $t_0$ (as they were not very important in the calculation of $L_{t_0}(g)$).

\subsection{Gradient calculation}
We optimize $D(\omega)$ with a gradient based algorithm and hence $\nabla D$
is needed. Let $i$ and $k$ be two distinct items $i\neq k$, then

\begin{eqnarray*}
\frac{\partial P(i|u,\omega)}{\partial \omega_k}&=&-\frac{\omega_iP(i|u)P(k|u)}{\left(\sum_{j\in I}\omega_jP(j|u)\right)^2}\\
&=&-P(i|u,\omega)\frac{P(k|u,\omega)}{\omega_k}.
\end{eqnarray*}

We have also
\begin{equation*}
  \label{eq:nabla:Pi:i}
\frac{\partial P(i|u,\omega)}{\partial \omega_i}=\frac{P(i|u,\omega)}{\omega_i}\left(1-P(i|u,\omega)\right),
\end{equation*}

and therefore for all $k$:

\begin{equation*}
  \label{eq:nabla:Pi:both}
  \frac{\partial P(i|u,\omega)}{\partial
  \omega_k}=\frac{P(k|u,\omega)}{\omega_k}\left(\delta_{ik}-P(i|u,\omega)\right).
\end{equation*}

We have implicitly assumed that the evaluation is based on independent draws, and therefore:
\begin{equation*}
  \label{eq:notation:Pik}
P(i,k|\omega)=\sum_{u}P(i|u,\omega)P(k|u,\omega)P(u).
\end{equation*}

Then
\begin{equation*}
  \label{eq:nabla:D}\notag
\frac{\partial D(\omega)}{\partial \omega_k}=\sum_{i}\frac{P_{t_0}(i)}{\omega_kP_{t_1}(i|\omega)}\left(P_{t_1}(i,k|\omega)-\delta_{ik}P_{t_1}(k|\omega)\right).
\end{equation*}

And in the particular case of uniform selection, \textit{i.e.} if $P(u) \sim \mathcal{U}(U)$ and $P(i|u) \sim \mathcal{U}(I_u)$, then:
\begin{eqnarray*}
 P(u) 	&=& \frac{1}{\#U}\\
 P(i|u) &=& \frac{1}{\#I_u} \cdot \mathds{1}_{i \in I_u}\\
 P(i|\omega) &=& \frac{1}{\#U} \cdot \sum_{u \in U_i }\frac{\omega_i}{\sum_{j \in I_u}\omega_j}\\
 P(i,k|\omega) &=& \frac{1}{\#U} \cdot \sum_{u \in U_i \cap U_k}\frac{\omega_i \omega_k}{(\sum_{j \in I_u}\omega_j)^2}
\end{eqnarray*}

\subsection{Complexity}\label{sec:complexity}
The value of $p$ coordinates of the gradient can be computed with a $\mathcal{O}(p\times n_{U \times I})$ complexity, where $n_{U\times I}$ is the number of couples $(u,i)$ wih $u \in U$ and $i \in I_u$ ($n_{U \times I} = \sum_{i \in I}\#U_i$). 

Indeed let us assume we have computed the beadjacency sparse matrix $B$ of the bipartite graph twice: once indexed by raws, and once indexed by columns. Such matrix can be got in $\mathcal{O}(n_{U \times I})$ and give access to every element in $\mathcal{O}(1)$. Then, in the particular case of uniform sampling it is possible to compute $P(i|\omega)$ for all $i\in I$ in $\mathcal{O}(n_{U\times I})$. 

Then if $\sum_{u\in U_i} \frac{\omega_i}{(\sum_{j \in I_u} \omega_j)^2}P(u)$ has been computed and saved for all $i$ (complexity in $\mathcal{O}(n_{u \times I})$), we have $P(i,k|\omega)$ in $\mathcal{O}(\#{U_i})$ for all $k$.

So, after having computed and saved the values of $P_{t_0}(i)$ and $P_{t_1}(i)$ for all $i$, the quantity $\frac{\partial D(\omega)}{\partial \omega_k}$ is a sum of $\#I$ elements computed in $\mathcal{O}(\#{U_i})$ and every coordinate of the gradient can be computed in $\mathcal{O}(\sum_{i \in I}U_i) = \mathcal{O}(n_{U \times I})$

\section{Illustration on constant algorithms}\label{sec:exper-eval-constant}

\subsection{Data and metrics}
We consider real world data extracted from Viadeo, where skills are attached to user's profile. The objective of the recommendation systems consists in suggesting new skills to users. The dataset contains 18294 users and 180 items (skills), leading to 117376 couples $(u,i)$. 

Both probabilities $P_t(u)$ and $P_t(i|u)$ are uniform, and the quality function $l$ is
given by $l(g_t(u_{-i}),i)=\mathds{1}_{i\in g_t(u_{-i})}$ where $g_t(u_{-i})$
is a set of 5 items. The quality of a recommendation algorithm, $L_t(g)$, is estimated via stochastic
sampling in order to simulate what could be done on a larger data set than the
one used for this illustration. We selected repeatedly 20 000 couples (user, item) (first we select a user $u$ uniformly, then an item according to $P_t(i|u,\omega)$).

\begin{figure}[htb]
  \centering
\includegraphics[width=5cm]{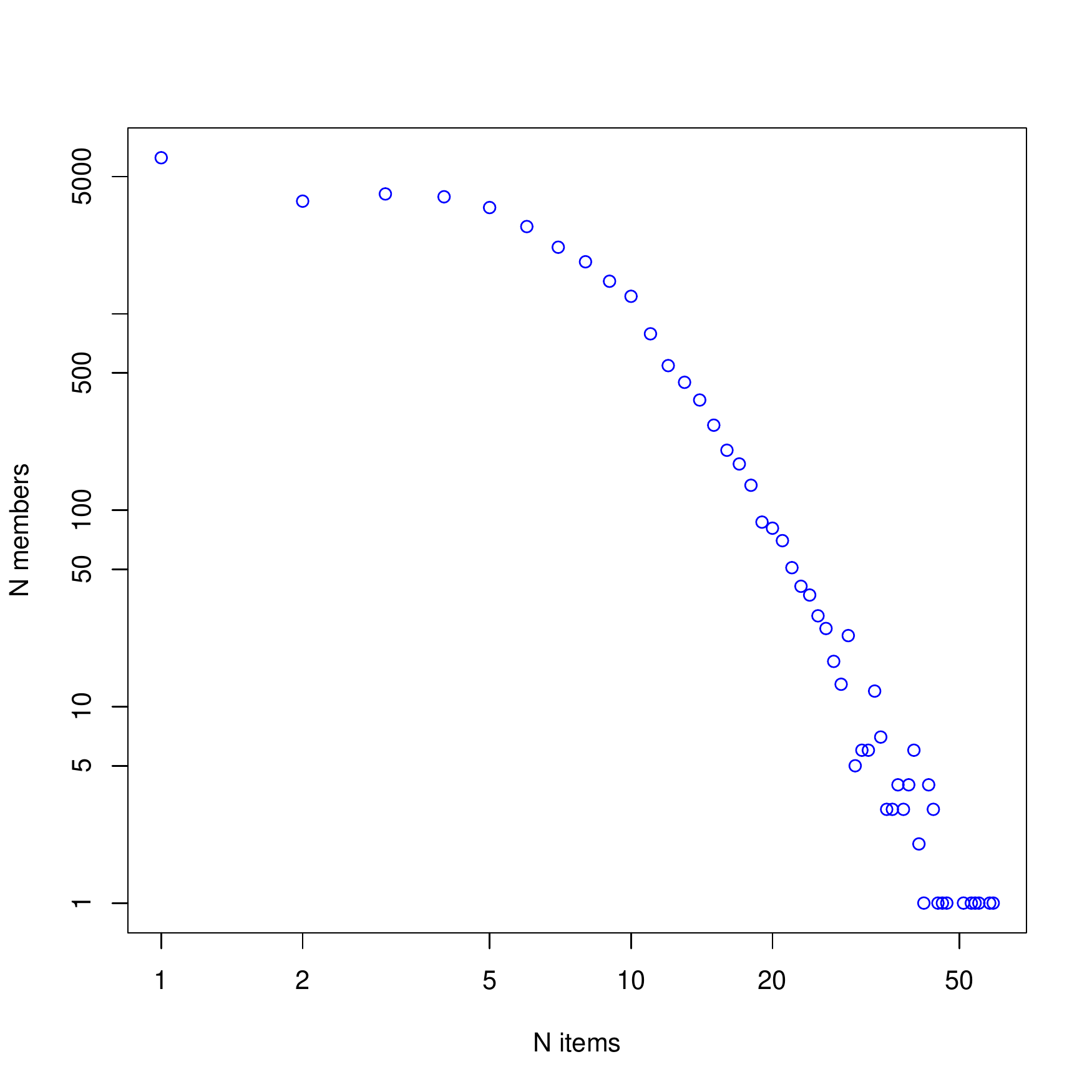}  
  \caption{Distribution of items per user}
  \label{fig:distribution}
\end{figure}

The recommendation setting is the one described
in Section \ref{sec:real-world-illustr}: users can attach skills to their
profile. Skills are recommended to the users in order to help them to build more
accurate and complete profiles. In this context, items are skills. The data
set used for the analysis contains 34 448 users and 35 741 items. The average
number of items per user is 5.33. The distribution of items per user follows
roughly a power law, as shown on Figure \ref{fig:distribution}. 

\subsection{Impact of previous recommendations campaigns}\label{sec:impactOnReco}
As described in section \ref{sec:offlineProcedure}, the offline evaluation of a recommendation algorithm can by computed using stochastic or exhaustive approach. Here we will describe the impact of previous recommendation campaigns on the offline evaluation score and compute the score of offline evaluation by stochastic sampling on the sample data extracted from Viadeo, what permits to mimic the results which could be computed on bigger datasets. We first demonstrate the effect of the bias on two constant recommendation
algorithms. The first one $g^1$ is modeled after the actual recommendation algorithm
used by Viadeo in the following sense: it recommends the five most recommended
items from $t=320$ to $t=480$. The second algorithm $g^2$ takes the opposite
approach by recommending the five most frequent items at time $t=300$ among
the items that were never recommended from $t=320$ to $t=480$. In a sense,
$g^1$ agrees with Viadeo's recommendation algorithm, while $g^2$ disagrees.

\begin{figure}[h]
  \centering
  \begin{subfigure}
  	\centering
	\includegraphics[width=4.5cm]{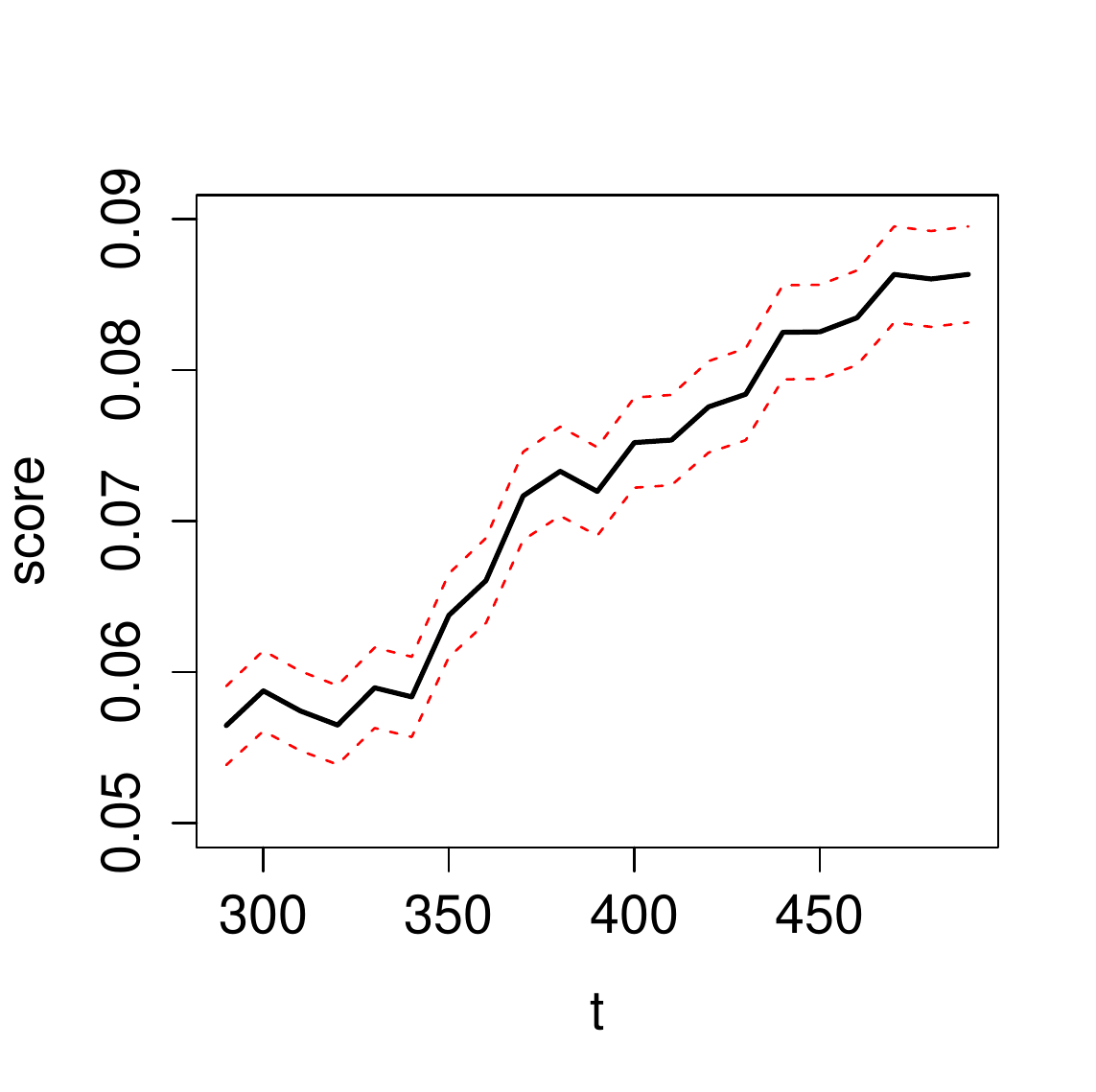}
	\label{fig:g:one}
  \end{subfigure}  
  \begin{subfigure}
 	\centering
	\includegraphics[width=4.5cm]{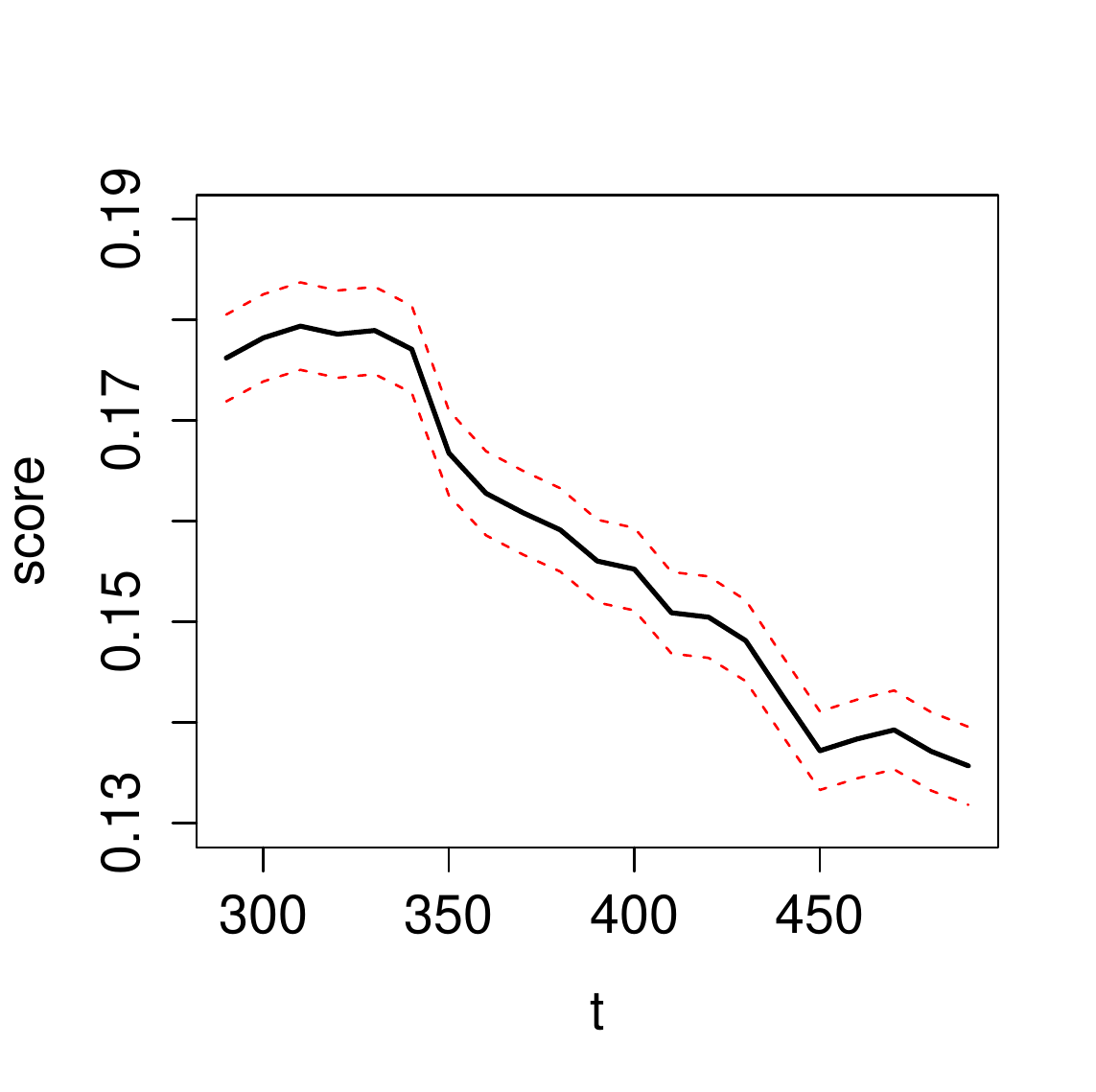}
	\label{fig:g:two}
  \end{subfigure}
  \caption{Evolution of $L_t(g^1)$ (left) and $L_t(g^2)$ (right) though time.}
  \label{fig:gconstant}
\end{figure}

For each couple of selected user and item $(u,i)$, the score given by the offline evaluation procedure of an algorithm $g$ is given by $l(g(u, i))$. For the experiments we have selected 30 000 couples $(u,i)$, where $u$ is a user chosen uniformly on $U$, and $i$ a skill chosen uniformly on $I_u$ (the set of skills associated to $u$).  We will consider the quality function given by $l(g(u, i)) = \mathds{1}_{i \in g(u, i)}$, where $g(u,i)$ represents the top fives items suggested by the algorithm $g$ after selecting the couple $(u,i)$. Figure \ref{fig:gconstant} shows the evolution of $L_t(g^1)$ and
$L_t(g^2)$ over time. As both algorithms are constant, it would be reasonable to
expect minimal variations of their offline evaluation scores. However in practice the estimated quality of $g^1$ increases by more than 25~\%, while the relative
decrease of $g^2$ reaches 33 \%.

As $l$ is a binary function, it can be considered as a Bernoulli random variable of parameter $p$, where $p$ corresponds to the expected probability that $l(g(u, i)) = 1$. Then, after $n_B$ simulations we have $n_B$ observations $\big(l(g(u_{(k)}, i_{(k)}))\big)_{k = 1, \dots, n_B}$ (where $u_{(k)}$ and $i_{(k)}$ corresponds to the user and item selected during the $k^{th}$ step of the offline evaluation procedure) and the maximum likelihood estimator of $p$ is given by 

\[ \widehat{p} = \sum_{k = 1}^{n_B} \frac{l(g(u_{(k)}, i_{(k)}))}{n_B} \]

Thus $\widehat{p}$ follows a binomial law which can be approximated by a gaussian random variable for $n_B$ big enough, and a $95\%$ confident interval for $\widehat{p}$ is classicaly given by 

\[ IC_{95\%}(\widehat{p}) = \bigg[ \widehat{p}  - 1.96 \sqrt{\frac{\widehat{p} (1-\widehat{p})}{n_B}} ; \widehat{p}  + 1.96 \sqrt{\frac{\widehat{p} (1-\widehat{p})}{n_B}}\bigg]\]

\subsection{Reducing the bias}

\begin{figure}[h]
  \centering
  \begin{subfigure}
  \centering
	\includegraphics[width=0.45\textwidth, keepaspectratio]{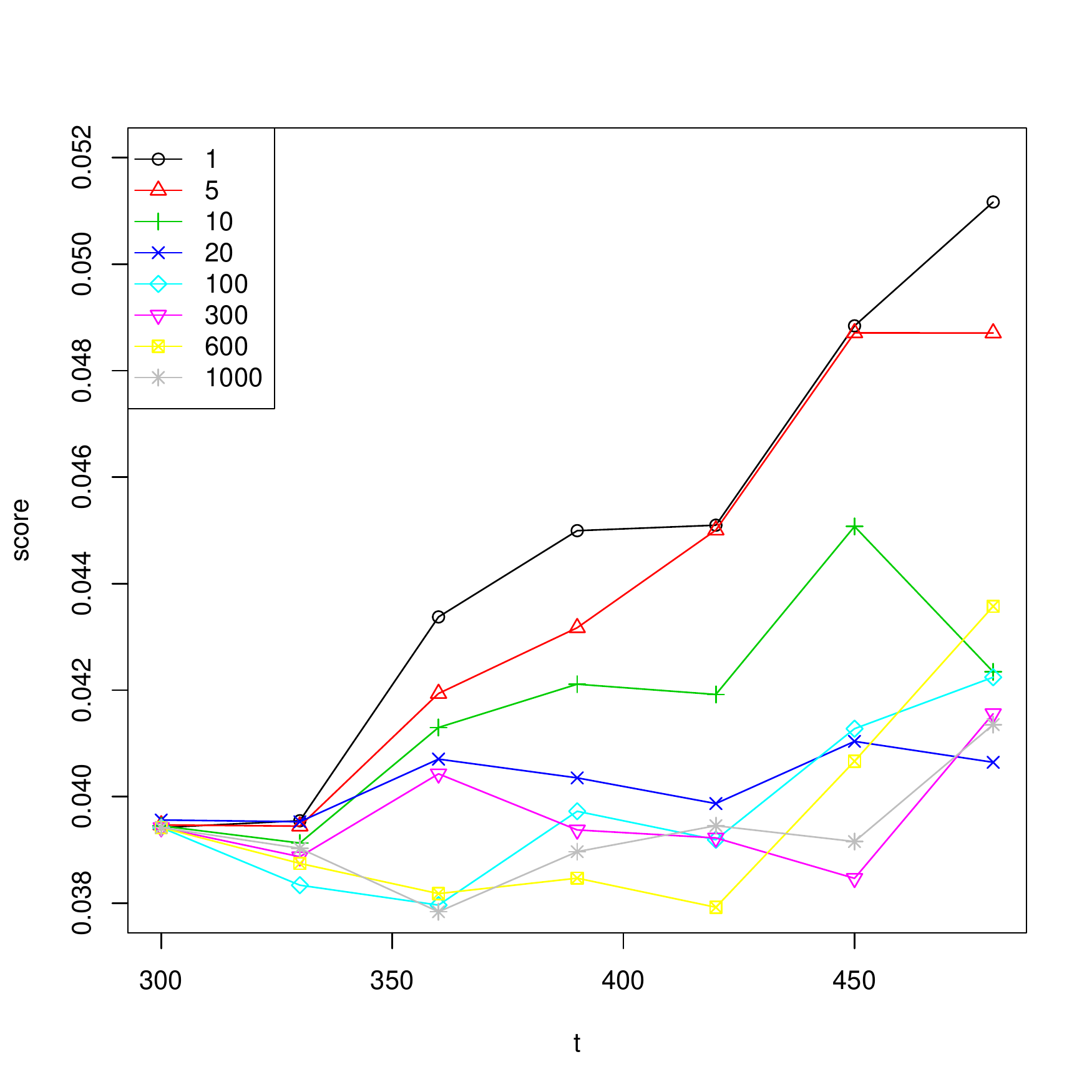}
  \end{subfigure}  
  \begin{subfigure}
  \centering
	\includegraphics[width=0.45\textwidth, keepaspectratio]{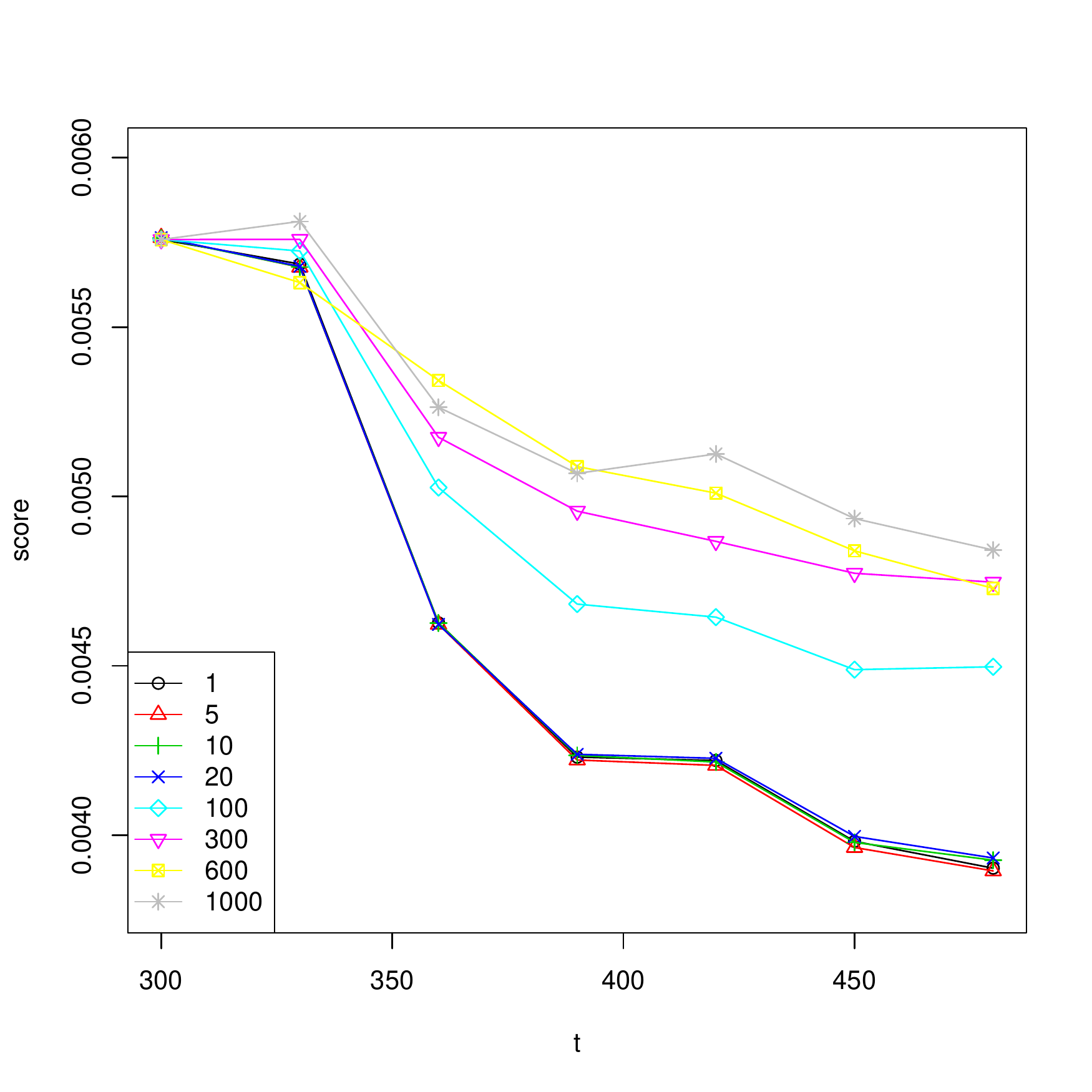}
  \end{subfigure}
  \caption{Evolution of $L_t(g^1)$ (left) and $L_t(g^2)$ (right) though time.}
  \label{fig:gconstantWeighted}
\end{figure}

We apply the strategy described in Section~\ref{sec:reduc-eval-bias} to
compute optimal weights at different instants and for several values of the
$p$ parameter. Results are summarized in Figure \ref{fig:gconstantWeighted}. 

The figures show clearly the stabilizing effects of the weighting strategy on
the scores of both algorithms. In the case of algorithm $g^1$, the
stabilisation is quite satisfactory with only $p=20$ active weights. This is
expected because $g^1$ agrees with Viadeo's recommendation algorithm and 
therefore recommends items for which probabilities $P_t(i)$ change a lot
over time. Those probabilities are exactly the ones that are corrected by the
weighting method.

The case of algorithm $g^2$ is less favorable, as no stabilisation occurs with
$p\leq 20$. This can be explained by the relative stability over time of the
probabilities of the items recommended by $g^2$ (indeed, those items are not
recommended during the period under study). Then the perceived reduction in
quality over time is a consequence of increased probabilities associated to other
items. Because those items are never recommended by $g^2$, they correspond to direct
recommendation failures. In order to stabilize $g^2$ evaluation, we need to
take into account weaker modifications of probabilities, which can
only be done by increasing $p$, as represented on figure \ref{fig:gconstant}.

Thus, the weighted offfline evaluation procedure reduces the bias for the very simple class of constant algorithms. In the next part we discuss the relevance of this procedure to reduce the offline evaluation bias on collaborative filtering algorithms.


\section{Experimentations on a collaborative filtering}\label{sec:exper-eval-cf}

\subsection{Collaborative filtering algorithms}\label{subsec:cf}
Collaborative filtering is a very popular class a recommendation algorithms which consists in computing recommendation to a user $u$ using the information available on other users, especially the ones similar to $u$. For example, a classical collaborative filtering consists in recommending the most frequent items among the ones associated to users having items in common with the user $u$.

More formally, let $B_{u}^t$ be the vector of items of user $u$ at time $t$ ($B_{u}^t  \in \{0,1\}^{\#I}$). Then $B_{u}(t)$ is a sparse vector as most of users are associated to only a few items, and corresponds to the $u^{th}$ column of the biadjacency matrix representing $\mathcal{D}_t$. The objective of collaborative filtering algorithms is to estimate $B_{u}^{t'}$ for $t'>t$ using the information known on other users. In this section we will discuss the efficiency of our method to reduce the offline evaluation bias on two different collaborative filtering algorithms:

\[ a) \widehat{B}_{u}^{t'} = \sum_{v \in U\backslash\{u\}} \frac{\langle B_{u}^{t}, B_{v}^{t}\rangle}{\sqrt{\|B_{u}^{t}\| \cdot \|B_{v}^{t}\|}}\cdot B_{v}^{t}\quad  b) \widehat{B}_{u}^{t'}(i) = \max_{j \in I_u(t)} \frac{\# (U_i \cap U_j)}{\#U_j} \]

The equation $a)$ is known as collaborative filtering with cosine similarity, whereas the equation $b)$ computes the proportion of users associated to item $i$ among the one associated to items possessed by $u$. Then we will note \emph{naive CF} (Collaborative Filtering) the algorithm $b)$.

Finally, the recommendation strategy consists in recommending the $k$ items having the highest values in $\widehat{B}_{u}^{t'}$.

\subsection{Results}
\begin{figure}[h]
\centering
\subfigure[cosine similarity]{\includegraphics[height=4cm]{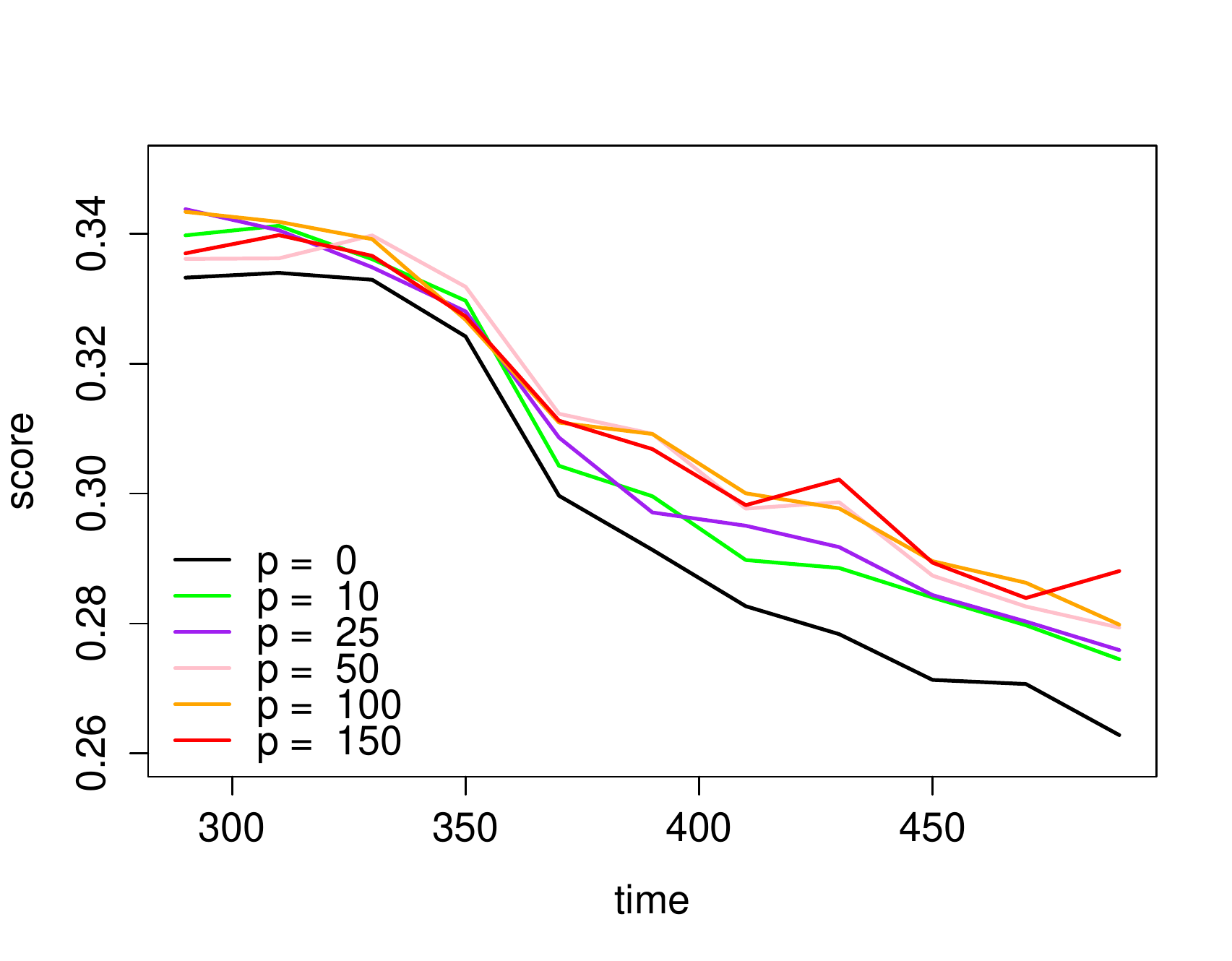}}
\subfigure[naive CF]{\includegraphics[height=4cm]{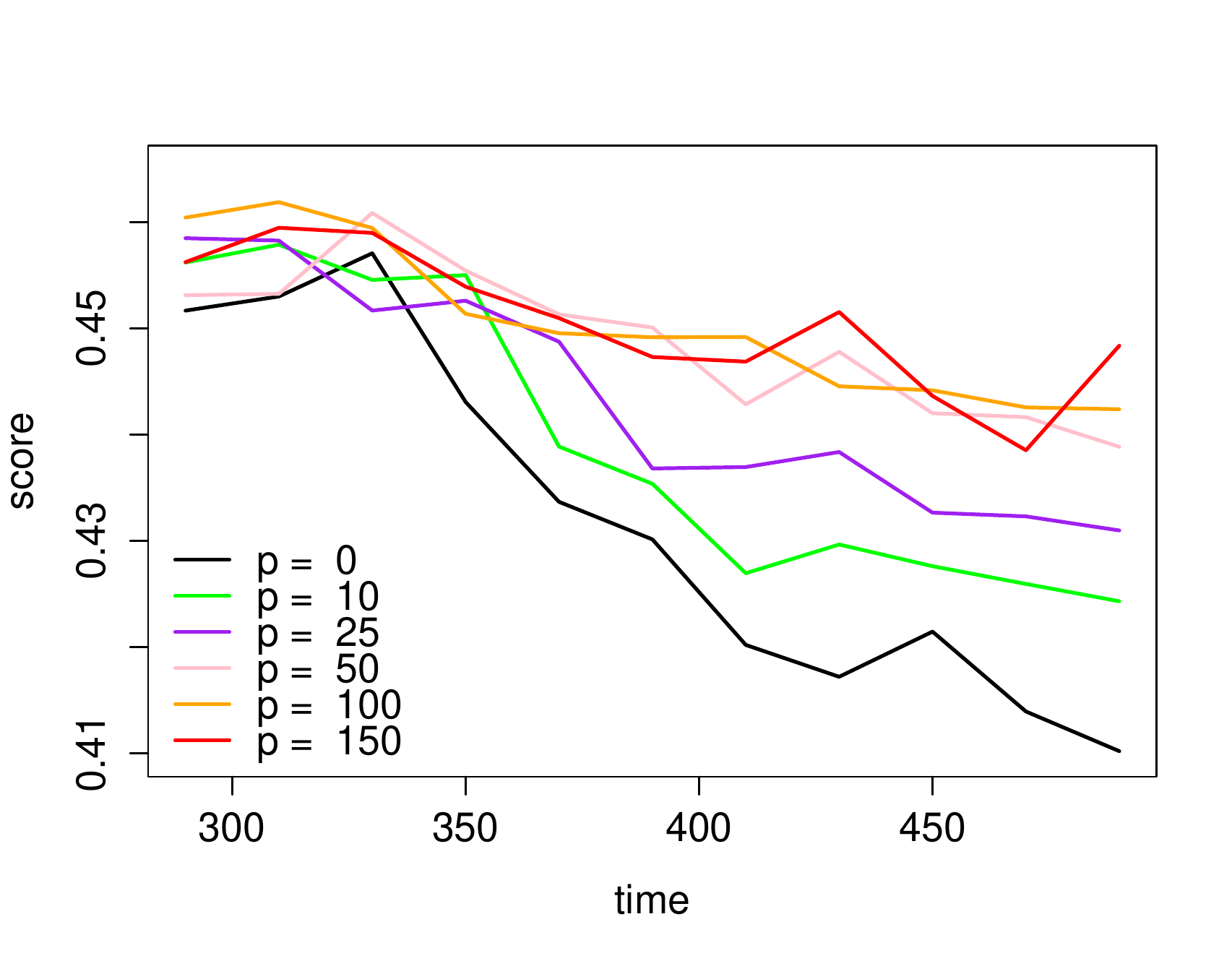}}
\caption{Results on the collaborative filtering with cosine similarity and naive CF, respectively defined by equation $a)$ and $b)$ in section \ref{subsec:cf}, for several values of $p$ (the number of weights optimized).}
\label{fig:cf}
\end{figure}

We apply the method described in Section~\ref{sec:reduc-eval-bias} to
compute optimal weights at different instants and for several values of the
parameter $p$. The collaborative filtering algorithms are the one presented in section~\ref{subsec:cf}. Results are summarized in figure \ref{fig:cf}. 

Once again the analysis is conducted on a 201 days period, from day 300 to day 500, where day 0
corresponds to the launch date of the skill feature and it is important to notice that two recommendation campaigns were conducted by Viadeo during this period at $t=330$ and $t=430$ respectively. As we can see on figure~\ref{fig:cf}, the scores strongly decrease after the first recommendation campaign ($t = 330$). Thus those campaigns have strongly biased the collected data, leading to a significant bias in the offline evaluation.

The figure~\ref{fig:cf} shows the influence of the value of $p$: the higher is $p$ the more weights are optimized and the more the bias is corrected. However, the efficiency of the recalibration depends on the algorithms. The results show that the weighting protocol permits to reduce the impact of recommendation campaigns on offline evaluation results as intended. However it does not lead to the stabilization of the score of collaborative filtering algorithms (while it lead to constant scores for constant algorithms). This can be explained by the nature of collaborative filtering: we can't expect the score to be constant for such an algorithm as it depends on the correlation between users, which have been modified by the recommendation campaigns. In others words the bias can be decompose in two parts: one depending on the probability selection of each item, and the second one depending on the structure of the data (the vertices in the bipartite graph representing the data). Indeed the structure of the graph $D_t$ has been modified because since recommendation campaigns have increased the density of the graph by adding new vertices from targeted users to recommended items.

\section{Conclusion}
Various factors influence historical data and bias the score obtained by classical offline evaluation strategy. Indeed, as recommendations influence users, a recommendation algorithm in production tends to be favored by offline evaluation. 

We have presented a new application of the item weighting strategy inspired by techniques designed for tackling the covariate shift problem. Whereas our previous results presented the efficiency of this method for constant algorithms, we have shown that this method also reduces the bias of more elaborate algorithms. However experiments on collaborative filtering shows that the bias can be decomposed in two part since previous recommendation campaigns change the probabilty selection of each item, but also modify the structure of the data. 

Experiments shows that our is efficient to reduce the first bias. Future works will invesgate the correction of the structural bias.


\begin{footnotesize}

\bibliographystyle{abbrv}

\bibliography{biblio}

\end{footnotesize}


\end{document}